\documentclass[showkeys,twocolumn,showpacs,preprintnumbers,amsmath,amssymb,pre]{revtex4}
\usepackage{Depken,graphicx,psfrag}
\bibstyle{apsrev}
\begin{document}

\title{Exact probability function for bulk density and current in the asymmetric exclusion process}
\author{Martin Depken and Robin Stinchcombe}
 \affiliation{University of Oxford, Department of Physics, Theoretical Physics\\ 1 Keble Road, Oxford, OX1 3NP, U.K.}
 \altaffiliation[MD's present address: ]{Instituut-Lorentz, Leiden University, P.O. Box 9506, 2300 RA Leiden, Netherlands\\ email: depken@lorentz.leidenuniv.nl}
\date{\today}
\begin{abstract}
We examine the asymmetric simple exclusion process with open boundaries, a paradigm of driven diffusive systems, having a nonequilibrium steady state transition. We provide a full derivation and expanded discussion and digression on results previously reported briefly in M. Depken and R. Stinchcombe, Phys. Rev. Lett. {\bf 93}, 040602, (2004). In particular we  derive an exact form for the joint probability function for the bulk density and current, both for finite systems, and also in the thermodynamic limit. The resulting distribution is non-Gaussian, and while the fluctuations in the current are continuous at the continuous phase transitions, the density fluctuations are discontinuous. The derivations are done by using the standard operator algebraic techniques, and by introducing  a modified version of the original operator algebra. As a byproduct of these considerations we also arrive at a novel and very simple way of calculating the normalization constant appearing in the standard treatment with the operator algebra.  Like the partition function in equilibrium systems, this normalization constant is shown to completely characterize the fluctuations, albeit in a very different manner.
\end{abstract}


\maketitle
\section{Introduction}
Since it is typically not feasible to describe macroscopic complex systems in terms of the exact dynamical evolution of their microscopic degrees of freedom, our knowledge of the collective properties of such systems has come mainly through using the methods of statistical mechanics. The systems are there described by a reduced set of variables (not necessarily in a direct correspondence to any physical observables) evolving according to stochastic dynamics, modeling the effect of the suppressed degrees of freedom. For equilibrium systems this approach has been extremely fruitful. On the other hand, recent insights into the properties of non-equilibrium steady-states (NESS) have come largely from specific microscopic studies (see below) and our general understanding of these steady states is still far inferior to that of equilibrium steady-states (ESS). 

On a phenomenological level, the distinction between ESS and NESS lies in that only the latter allows for net probability currents through the state-space (in equilibrium detailed balance excludes this). The currents are set up and maintained by contacts with multiple reservoirs at different potentials (e.g. different temperature or chemical potential), and/or by the presence of non-conservative bulk forces. We will refer to the former case as a boundary driven system, and to the latter as a bulk driven system. On a  formal level the difference between a ESS and a NESS lies in that for ESS the probability weights of different configurations are governed by detailed balance, while there is no such generic rule for NESS. Consequently the probabilistic weights of the configurations in a NESS are not {\it a. priori.} known (contrast the Gibbs weights of ESS), and the lack of detailed balance allows for net-currents through the configuration space. Apart from this, there is also the fact that ESS with short ranged interactions generically (i.e. in the absence of a spontaneously broken continuous symmetry, and not at a continuous phase transition) have finite correlation lengths. Therefor one can adopt an ensemble approach to fluctuations for many-particle ESS~\cite{Landau80}. For a NESS on the other hand, there can exist long-range correlations even when the system is not externally tuned to be at a phase transition~\cite{Garrido90}. This observation has lead to the ideas of self-organized criticality, whereby the system generates infinite correlation lengths by self-adjusting its' state to a critical state~\cite{Bak87}. By now there are also examples of non-critical NESS, which still show infinite correlation lengths~\cite{Zia95,Bergersen95,Alexander98}  or (as in the system studied below) where boundary driving is felt throughout the system .

Faced with these complications one does wisely in utilizing methods that have proved their worth elsewhere. One highly effective tool within the study of ESS is the use of specific simplified and abstracted models. Among these are the classical- and quantum-magnetic models, the Ising model, and the Heisenberg models to name only two. These models were initially only hoped to give a qualitative description of the physical systems. They retain only some of the essential ingredients, while leaving out enough of the details for progress to be possible. With the advent of renormalization group arguments~\cite{Cardy96}, and the implied universality, these models are now known to also give quantitative predictions, and the study of simple models is now a fundamental part of the understanding of ESS. 

Led by this, and the fact that universality seems to be a generic feature also of the long wavelength and low frequency degrees of freedom of NESS~\cite{Halpin-Healy95}, much of the effort within the study on NESS has focused on particular simple models. Here we study a well know simple $1+1D$ model of a driven diffusive system, the asymmetric simple exclusion process (ASEP). It belongs to the class of stochastic interacting particles models~\cite{Liggett85}, and the particular version we choose to study is driven both from the boundary and the bulk. This model is non-trivial, displaying macroscopic collective behavior in the form of both continuous and discontinuous boundary-induced phase transitions. It is yet simple enough to be integrable~\cite{Derrida92,Derrida93,Sasamoto99,Blythe99}. It is of additional interest since it maps onto certain growth models~\cite{Halpin-Healy95}, it models traffic flow~\cite{Schreckenberg98}, and it is believed to describe the large scale dynamics of the noisy burgers equation~\cite{Burgers74,Gwa92,Stinchcombe01} and the KPZ equation~\cite{Kardar86,Stinchcombe01}. There has been much progress in the analytical treatment of the ASEP, giving rise to a host of exact results describing its' steady state properties~\cite{Derrida92,Derrida93,Sasamoto99,Blythe99,Ferrari94,Lee99,Derrida98,Derrida02}. The probability measure of this model (and of certain other one-dimensional models (see e.g. section 6 of~\cite{Stinchcombe01})) turns out to be a generalization of the trivial product measure,  where instead of a measure built up of commuting $c$-numbers at each spatial position, the measure is constructed by assigning non-commuting operators to each spatial position (for further details see below). Even though the fact that the system maintains a current is ultimately what sets it apart from an equilibrium system, the results so far concerning the currents are mainly for systems with periodic boundaries or infinite geometries with special initial conditions~\cite{Ferrari94,Lee99,Praehofer02,Derrida03}. 

In this paper we expand the discussion and digress on what was briefly reported in~\cite{Depken04}.  We will derive the exact joint probability distribution for the system averaged current and density, as well as their asymptotic form in the thermodynamic limit. This is a step closer to the achievements of statistical mechanics for ESS, and comparable to summing the partition function; but it is system-specific. 

The paper is organized as follows; Section~\ref{sec:def} gives the definition of the model, recapitulates a few known results, as well as giving a brief introduction to the operator algebraic approach. This is followed by Section~\ref{sec:mac} where we introduce a novel way of calculating the normalization factor (first introduced in Section~\ref{sec:opalg}), and further introduce a ``relaxed'' operator algebra, with which help we are able to calculate the exact form of the density-current probability function. This is complemented by the study of the thermodynamic limit. In Section~\ref{sec:KPZ} we translate the results into the language of the KPZ-equation. The details of the calculations of the preceding sections are recorded in the appendices.

\section{Model Definition and some known results}
\label{sec:def}
\begin{figure}[htp]
\psfrag{a}{$\alpha$}
\psfrag{b}{$\beta$}
\psfrag{c}{$1$}
\psfrag{d}[cl]{$1\ldots$}
\psfrag{e}[c]{$\ldots l \ldots$}
\psfrag{f}[cr]{$\ldots L$}
\includegraphics[width=.9\columnwidth]{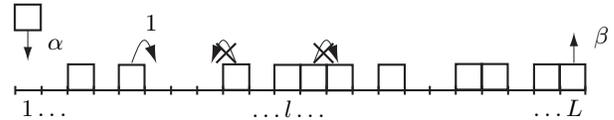}
\caption{\label{fig:mod} Illustration of the dynamical rules of the ASEP}
\end{figure}

We consider the totally asymmetric exclusion process (ASEP) on a finite chain of size $L$ with open boundaries. The site label $l$ runs from left to right (see Figure~\ref{fig:mod}). 
Each site on the lattice can be occupied by no more than one particle, $n_l\in\{0,1\}$. Given that the right neighboring site of an occupied site is empty, the occupying particle will jump to the empty site with rate $1$. If the first site  on the lattice is unoccupied, particles are injected at this boundary with rate $\alpha$. Further given that we have a particle at the last site  of the lattice, it is ejected with the probability rate $\beta$. No further transitions are allowed. We will here limit our considerations to the case where we can view the boundary rates as deriving from particle reservoirs. We therefore take $0<\alpha=\rho_{{\rm left}}<1$ and $0<\beta=1-\rho_{{\rm right}}<1$, where $\rho_{{\rm left}}$ and $\rho_{{\rm right}}$ are the particle densities of the reservoirs.
\subsection{The phase diagram}
\label{sec:phase}
\begin{figure}[htp]
\hspace{-1cm}\includegraphics[width=.8\columnwidth]{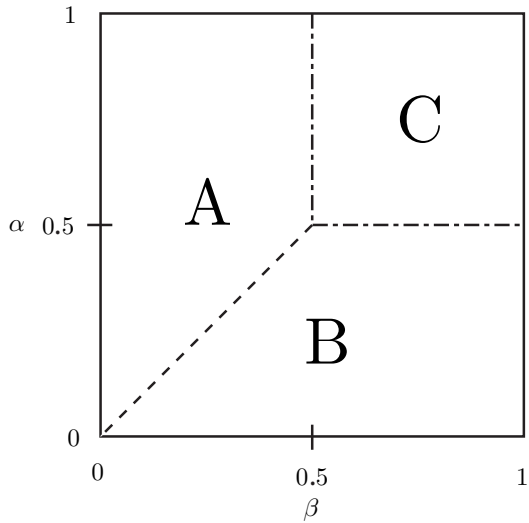}
\caption{\label{fig:pd} Phase diagram of the one dimensional exclusion process. The dashed  and dash-dotted lines indicate respectively first order and continuous transition lines.}
\end{figure}

 This model has been exactly solved~\cite{Derrida93}  in the sense that the steady-state probability of any given microscopic configuration can (in principle) be calculated by applying a given set of algebraic rules (described below). This though is far from what we are used to refer to as a solution in equilibrium statistical mechanics. There we are in a better position already from the start in that the configurational weights are explicitly given, while for this model they are given only in terms of algebraic rules (or matrix multiplication in case we have a finite matrix representation of the operator algebra (see below)). In equilibrium it is normally the successful summation of the partition function that is considered a solution of the problem. Even so, the algebraic rules yield a wealth of information about the system. Especially they can be used to deduce the precise form of the phase-diagram~\cite{Derrida93}. It consists of three parts A, B, and C, as given in Figure~\ref{fig:pd}. 
In part A, the low-current, high-density phase, the average bulk profile, $\rho_l=\lv n_l\rv$, is flat in the bulk with $\rho_{l}=\rho_{{\rm right}}=1-\beta$, and the current is $j_l=\beta(1-\beta)$. The average bulk profile is completely dictated by the right hand side reservoir. The bulk profile connects to the value of the left reservoir through an exponential decay, with some characteristic length. The situation is reversed in B, the low-current, low-density phase, in that the bulk density profile is dictated by the left reservoir, with $\rho_l=\rho_{{\rm left}}=\alpha$, and $j_l=\alpha(1-\alpha)$. The decay of the density profile to the right boundary value $\rho_{{\rm right}}=1-\beta$ is also here exponential. Separating the two phases is a phase transition at which the typical bulk profile develops a kink, taking the density from the value of the left reservoir to the value of the right one. The kink is of finite extension, and equally likely to be situated anywhere throughout the bulk. When passing over this transition line the density is discontinuous, and the border between A and B is thus a first-order non-equilibrium phase-transition. Such a boundary induce phase transition has no counterpart in equilibrium systems where boundaries are assumed irrelevant. The remaining region, region C, is called the maximal-current phase. Here the bulk properties are set by the bulk drive, i.e. in this phase both the injection and ejection rate, are high enough for the system only to be limited by the transition rates in the bulk. The bulk profile is given by $\rho_l=1/2$ and $j_l=1/4$. In this phase the decays from the boundaries are algebraic.

 A nice heuristic argument for the form of the boundary decays has recently been presented in~\cite{Ha03}. The transitions from  A or B to C induces a jump in neither the average density nor current, and thus these transitions are continuous. As we will argue later, they are in fact second order. The point where all regions meet is called the critical point.  These results can all be derived by utilizing the matrix algebra outlined below. Further results containing information about the time evolution of the system can be derived by Bethe-ansatz methods~\cite{Dhar87,Gwa92,Kim95,Derrida98}.

\subsection{Operator algebra}
\label{sec:opalg}
We here outline the operator algebra as given in~\cite{Derrida93} since we will be using and generalising this. The starting point is to represent any microscopic configurations in terms of a string of non-commuting operators $\bm  D$ and $\bm  E$, corresponding to a particle and a hole respectively. It can then be shown that the steady-state probability function can be written in terms of this operator string and two auxiliary vectors, $\lv \alpha |$ and $|\beta\rv$, according to
\be{eq:ssm}
  P_{\rm ss}(\{n_l\})= (Z_L^{\alpha\beta})^{-1}\lv \alpha |\bm  X(n_1)\bm  X(n_2)\cdots\bm  X(n_L)|\beta\rv.
\ee
Here the operator $\bm  X(n_l)$ equals $\bm  D$ if there is a particle at site $l$ ($n_l=1$), and $\bm  E$ if site $l$ is unoccupied ($n_l=0$). The vectors $\lv \alpha|$ and $|\beta\rv$ describe the properties of an uncorrelated particle reservoir. The state independent factor  $Z_L^{\alpha\beta}=\lv \alpha |\bm  (\bm D+\bm E)^L|\beta\rv$ ensures the proper normalization. For~(\ref{eq:ssm}) to hold true, the operators and  vectors must further satisfy the algebraic rules
\be{eq:opalg}
  \bm  D\bm  E=\bm  E+\bm  D\Def\bm C, \,\,\,\, \lv \alpha |\bm  E=\frac{1}{\alpha}\lv \alpha |, \,\,\,\, \bm  D |\beta\rv=\frac{1}{\beta}|\beta\rv,
\ee
and we may take the normalizations of the vectors $|\alpha\rv$ and $\lv \beta|$ to be such that $\lv \alpha |\beta\rv=1$. The algebraic rules~(\ref{eq:opalg}) are now all that is needed to calculate $P^{\alpha \beta}_{\rm ss}(\{n_l\})$, resulting in a polynomial of degree $L$ in $1/\alpha$ and $1/\beta$. An alternative approach is to look for matrix representations of the algebra~(\ref{eq:opalg}), and then use these to calculate the microscopic probabilities. This justifies calling these states matrix-product states. Even though we now have a way of getting the steady-state weight of any specific configuration, actually calculating this number becomes increasingly hard as one considers larger and larger systems. Thus one wishes to extract general information directly from the algebraic rules, without explicitly calculating the microscopic weights. This is done for both average density and current in~\cite{Derrida93}, and we here just recall that it is in general very easy to write down the desired quantities in terms of the above defined operators
$$
\begin{array}{rl}
  \rho_l&\Def\lv n_l\rv =\lv \alpha|\bm  C^{l-1}  \bm  D \bm  C^{L-l}|\beta\rv/Z^{\alpha\beta}_{L}\\
  j_l&\Def\lv n_l(1-n_{l+1})\rv\\
  &=\lv \alpha |\bm  C^{l-1}\bm  D\bm  E \bm  C^{L-l-1}|\beta\rv/Z^{\alpha\beta}_{L}=Z^{\alpha\beta}_{L-1}/Z^{\alpha\beta}_{L}\\
  C(l,m)&\Def\lv n_l n_m\rv\\
& =\lv \alpha |\bm  C^{l-1}\bm  D\bm  C^{m-l-1}\bm  D \bm  C^{L-m}|\beta\rv/Z^{\alpha\beta}_{L}.
\end{array}
$$
The first two can be calculated asymptotically for large $L$ and fixed $x=n/L$~\cite{Derrida93}, giving the phase diagram discussed above. Many of the results for these systems have been derived through finding a matrix representation of the operator algebra. If there exists a finite sized matrix representation, then the above form of the correlation functions show that the inverse correlation length is simply proportional to the highest eigenvalue of the matrix representation of $\bm  C$. Thus, only when no finite representation can be found is it possible to have an infinite correlation length, and algebraic decay of correlations.
\section{Macroscopic description}
\label{sec:mac}
As is the case for ESS, the transition from a detailed microscopic knowledge about the weights of each configuration to information concerning macroscopic properties of the system is in general highly non-trivial (c.f. summing the partition function). Here though we are faced with one additional problem in that we do not have the microscopic weights explicitly, but only a set of algebraic rules~(\ref{eq:opalg}), or as a string of possibly infinite matrices. Thus, as mentioned above, we need to extract macroscopic quantities directly through using the algebraic rules. One macroscopic quantity that has been likened to a partition function of this non-equilibrium system is the normalizing constant $Z_L^{\alpha\beta}$~\cite{Blythe02}. Though it lacks the power of yielding moments through differentiation, it generates a form of  Lee-Yang theory of these non-equilibrium phase transitions. Through the considerations in this section we will further see that the normalizing constant still plays an instrumental role in determining the complete statistics of the bulk densities and currents in this non-equilibrium system.
\subsection{The generating function for $Z_L^{\alpha\beta}$}
As a warm up to what will follow, and since the normalization constant plays a central role in later developments, we here present a novel and very simple way of calculating $Z^{\alpha\beta}_{L}$. This is done through calculating the generating functional of the normalization constant, an approach recently used in~\cite{Depken03,Blythe04}. We define the generating functional as
\be{eq:gz}
  G^{\alpha\beta}(\mu)=\sum\limits_{L=0}^X \mu^L Z_L^{\alpha\beta}=\lv \alpha|\left[\frac{1}{1-\mu \bm  C}\right]_X|\beta\rv,
\ee
where for an arbitrary operator $\bm  X$ we have defined
\be{eq:angdef}
  \left[\frac{1}{1-\opa  X}\right]_X\Def\sum\limits_{L=0}^X \bm  X^L.
\ee
Here $X$ is some finite integer, and it immediately follows that
\be{eq:inv}
  (1-\mu \bm  X)\left[\frac{1}{1-\mu\bm  X}\right]_X=1+\O(\mu^{X+1}).
\ee
From the operator algebra~(\ref{eq:opalg}) we have
$$
  (1-\mu \bm  D)(1-\mu \bm  E)=1-\mu(1-\mu)\bm  C,
$$
and inverting this in the sense of~(\ref{eq:inv}) we get
$$
  \left[\frac{1}{1-\mu(1-\mu)\bm  C}\right]_X=\left[\frac{1}{1-\mu \bm  E}\right]_X\left[\frac{1}{1-\mu \bm  D}\right]_X+\O(\mu^{X+1}).
$$
In the above we note that all $\bm  E$'s are to the left of all $\bm  D$'s. Since $\lv\alpha|$ and $|\beta \rv$ by definition are respective eigenvectors of these operators, we have
$$
  G^{\alpha\beta}\bm(\mu(1-\mu)\bm)=\left[\frac{1}{1-\mu/\alpha}\right]_X\left[\frac{1}{1-\mu/\beta}\right]_X+\O(\mu^{X+1}).
$$
In the limit $X\rightarrow\infty$, the above expression has a convergence radius of $\min(\alpha,\beta)>0$. Thus as long as we are within this we can take this limit and write
$$
  G^{\alpha\beta}\bm (\mu(1-\mu)\bm)=\frac{\alpha}{\alpha-\mu}\frac{\beta}{\beta-\mu}.
$$
There now exists an open region around the origin where this can be rewritten as
$$
  G^{\alpha\beta}(\mu)=\frac{2\alpha}{2\alpha-1+\sqrt{1-4\mu}}\frac{2\beta}{2\beta-1+\sqrt{1-4\mu}}.
$$
The expression can be analytically continued to all $\mu$, and is inverted through
\be{eq:gen}
  Z^{\alpha\beta}_L=\oint_{C_\mu}\frac{\d \mu}{2\pi\imath}\frac{G^{\alpha\beta}(\mu)}{\mu^{L+1}},
\ee
where $C_{\mu}$ encircles only the pole at the origin, and does so once in the positive direction. This can be used to derive the same finite size form of $Z_L^{\alpha\beta}$ as given in~\cite{Derrida93},
\be{eq:norm}
  Z^{\alpha\beta}_L=\sum\limits_{l=1}^{L} A_{L,l}\sum\limits_{k=0}^{l}\frac{1}{\alpha^k\beta^{l-k}},\quad A_{L,l}=\frac{l(2L-l-1)!}{L!(L-l)!}.
\ee
One can also perform a  large system asymptotic analysis on~(\ref{eq:gen}) using steepest-descent methods, which again yields the same results as in~\cite{Derrida93},
\begin{eqnarray}
  1/2<\alpha < \beta:  &Z^{\alpha\beta}_L\sim&\frac{4^{L+1}\alpha\beta}{L^{3/2}\sqrt{\pi}}\frac{\alpha+\beta-1}{(2\alpha-1)^2(2\beta-1)^2},\label{eq:zh}\\
 \alpha<1/2,\beta: &Z^{\alpha\beta}_L\sim&\frac{\beta(1-2\alpha)}{(\beta-\alpha)(1-\alpha)}\frac{1}{\alpha^L(1-\alpha)^L},\nonumber\\
 \alpha=\beta<1/2:  &Z^{\alpha\beta}_L\sim&\frac{(1-2\alpha)^2}{(1-\alpha)^2}\frac{L}{\alpha^L(1-\alpha)^L}\label{eq:zl}.
\end{eqnarray}
The asymptotic forms for $\alpha>\beta$ can be obtained through realizing that the system exhibits a particle hole symmetry. That is, instead of focusing on the particles we might just as well consider the holes as evolving with exactly the same dynamics, but with the injection and ejection rates exchanged. Thus we can directly get the result for $\alpha>\beta$ by letting $\alpha\rightarrow \beta$ and $\beta\rightarrow\alpha$ in the above.
\subsection{The bulk current-density probability function}
Though the normalization constant considered above has some of the feature of the equilibrium partition function, it does not in it's present applications tell us much about the moments of the two natural observables of the system, the density and current. Thus we here concentrate on the derivation of the exact  joint probability function for the average density and current throughout the bulk. Through this we will see how the normalization constant also here tells us about fluctuations, albeit in a manner very different from that of equilibrium statistical mechanics. This is done for any system size, and later the thermodynamic limit is also considered. 

First we define the total activity within the system as the number of bulk bonds that can facilitate a transition of a particle in the immediate future, i.e. the total effective bulk transition rate. The bulk current is then defined as the activity divided by the system size. For any given state the activity equals the number of pairs of neighboring sites that  have a particle to the left and a hole to the right. To get a handle on the activity, $J$, of a microscopic configuration of $N$ particles we choose to represent such a configuration by a sequence of $J$ objects of the form $\bm D^{p_j}\bm E^{h_j}$, $p_j,h_j\ge 1$, possibly padded with $\bm E$'s to the left and $\bm D$'s to the right. Each of these objects contains what corresponds to an active bond, and using these objects we can write any microscopic steady state measure as
\begin{multline*}
P_{{\rm ss}}(\{p_j,h_j\})\\
= (Z_L^{\alpha\beta})^{-1}\lv \alpha |\bm  E^{h_0} (\bm D^{p_1}\bm E^{h_1})\cdots (\bm D^{p_J}\bm E^{h_J})\bm D^{p_0}|\beta\rv,
\end{multline*}
by appropriately choosing the numbers $\{p_j,h_j\}$ and $J$. It further follows that the above expression is unique if $h_0,p_0\ge 0$, and the rest satisfy $h_j,p_j\ge 1$. We can now in principle calculate the joint probability distribution for $N$ and $J$ by summing the above over all $h_j$'s and $p_j$'s consistent with a specific number of particles ($\sum_{j=0}^J p_j=N$) and a given system size ($N+\sum_{j=0}^J h_j=L$). Choosing to enforce these constraints with contour integral representations of the Kronecker delta, the expression for the joint particle-activity probability function can be written as
\begin{widetext}
\begin{multline}
\label{eq:obsprob}
  P^{\alpha\beta}_L(N,J)=\sum\limits_{p_0,h_0=0}^{L}\sum\limits_{p_{l},h_{l}=1, \, l\ge 1}^{N,L-N}P_{{\rm ss}}(\{p_j,h_j\})\delta_{\sum p_j,N}\delta_{\sum h_j,L-N}\\
=\frac{\alpha\beta}{Z_L}\oint\limits_{C_{z},C_{\bar z}} \frac{\d z\d \bar z}{(2\pi\imath)^2}\frac{1}{z^{N+1-J}\bar z^{L-N-J+1}}\frac{1}{(z-\beta)(\bar z-\alpha)}\lv \alpha |\left(\left[\frac{1}{1-z\bm  D}\right]_{N-1} \bm  D \bm  E\left[\frac{1}{1-\bar z \bm  E}\right]_{L-N-1}\right)^J
|\beta\rv.
\end{multline}
\end{widetext}
Here $C_z$ ($C_{\bar z}$) is a directed contour that encircle the pole at the origin of the complex $z$ ($\bar z$) plane once in the positive direction, with $|z|<\beta$ ($|\bar z|<\alpha$).  The first step toward explicitly calculating~(\ref{eq:obsprob}) is through considering the properties of involved operator product. Surprisingly  one can show (see Appendix~\ref{app:dual}) that a slight modification of the above operators
\be{eq:newop}
\begin{array}{rl}
  \bm D'&\Def[1-(z+\bar z)]\left[\frac{1}{1-z\bm  D}\right]_{N-1}\bm  D,\\
  \bm E'&\Def[1-(z+\bar z)]\left[\frac{1}{1-\bar z\bm  E}\right]_{L-N-1}\bm  E,
\end{array}
\ee
satisfy the ``relaxed''operator algebra
\be{eq:newopalg}
  \bm D'\bm E'=\bm D'+\bm E'+\O(z^{N},\bar z^{L-N}).
\ee
The new relaxed eigenvectors and eigenvalues are simply given by
$$
  \bm D'|\beta\rv =|\beta\rv\frac{1}{\beta'}+\O(z^N), \quad  \lv \alpha |\bm E'=\frac{1}{\alpha'}\lv \alpha |+\O(\bar z^{L-N}),
$$
with the eigenvalues defined as
\be{eq:eig}
\alpha'\Def\frac{\alpha-\bar z}{1-(z+\bar z)}, \quad
\beta'\Def\frac{\beta-z}{1-(z+\bar z)}.
\ee
The fact that these eigenvalues are complex is of no concern since we consider only finite polynomials in the inverse eigenvalues. Any result is thus uniquely extendable into the complex plane through analytic continuation. We can rewrite~(\ref{eq:obsprob}) in terms of the primed operators, and start using the relaxed operator algebra to transform the expression.  The result of any such manipulation would, according to the above, be the same up to terms of order $z^N$ and $\bar z^{L-N}$, as if the operator algebra would have been exact. Terms of this order have no effect under the contour integral in~(\ref{eq:obsprob})  since the poles at the origins are of order equal  to or lower than $N$ and $L-N$ respectively. (The case for $J=0$ is trivial.) Thus, using the new algebra to perform any manipulation within~(\ref{eq:obsprob}) is equivalent to using the exact algebra. Therefore we can write
\begin{multline}\label{eq:prob1}
   P^{\alpha\beta}_L(N,J)=\frac{\alpha\beta}{Z_L^{\alpha\beta}}\oint\limits_{C_{z},C_{\bar z}} \frac{\d z\d \bar z}{(2\pi\imath)^2}\frac{1}{z^{N+1-J}\bar z^{L-N+1-J}}\\ 
\times\frac{Z_J^{\alpha'\beta'}}{(z-\beta)(\bar z-\alpha)[1-(z+\bar z)]^{2J}}.
\end{multline}
This expression is the main result of this work, and since all quantities in it are known exactly, it yields both the exact finite system size form of $P^{\alpha\beta}_L(N,J)$, as well as the asymptotic form in the large system size limit. Using the above expression it is further easy to derive a similar form for the joint generating functional of the density and current. This is outlined in Appendix~\ref{app:gen}, while we  go on here and present exact and asymptotic results for the probability function.
\subsection{Exact results for finite systems}
\begin{figure}[htp]
  \includegraphics[width=.45\textwidth]{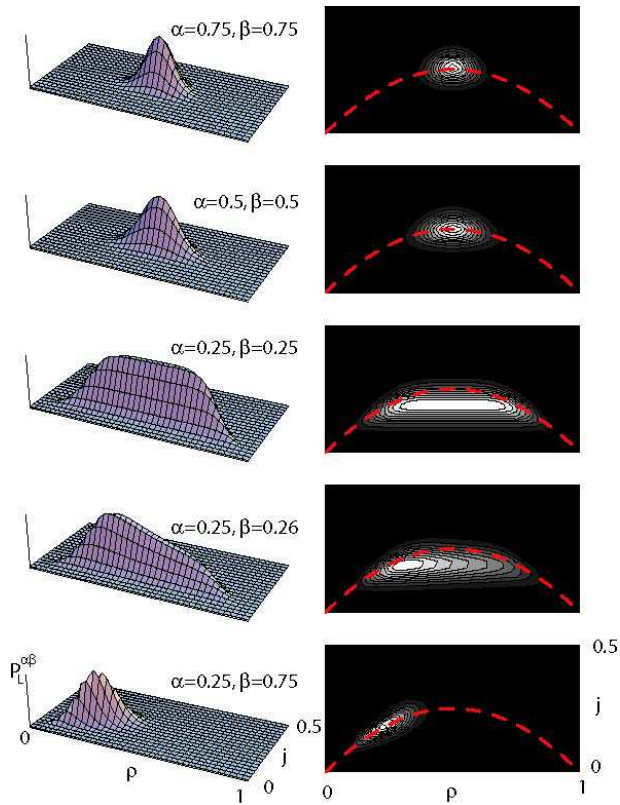}
\psfrag{=}{x}
\caption{\label{fig:Exact} (Color online) Each row contains a surface and a contour plot of the exact joint probability distribution for the values of $\alpha$ and $\beta$ indicated, and with $\rho=N/L$ and $j=J/L$. The first three rows illustrate the behavior of the probability distribution as the system goes along the line of $\alpha=\beta$ through the critical point at $\alpha=\beta=0.5$, while the last three graphs illustrate the behavior as the system goes through the first-order transition at $\alpha=\beta=0.25$.  Overlaid in the contour plots (dashed line) is the curve $j=\rho(1-\rho)$ which defines the set of possible asymptotic average values of $\rho$ and $j$ throughout the system's  different phases (not at the first order transition line). The system size is $L=40$.}
\end{figure}

 The integral in~(\ref{eq:prob1}) is easily calculated with the help Cauchy's integral theorem. All we need to do is to calculate the coefficient of the term proportional to $(z\bar z)^{-1}$ in the Laurent-series expansion of the integrand in~(\ref{eq:prob1}). For the special case $J=0$  we have $Z_J^{\alpha'\beta'}=1$, and thus
$$
  P^{\alpha\beta}_L(N,0)=\frac{1}{Z_L^{\alpha\beta}}\lp1/\beta\rp^N\lp1/\alpha\rp^{L-N}.
$$
This is obviously correct since the inactive state must have $L-N$ empty sites followed by $N$ filled sites. In Appendix~\ref{app:exact} we consider the case $J\ge 1$. The result is 
\begin{widetext}
\be{eq:K}
  P^{\alpha\beta}_L(N,J)=\frac{\alpha\beta}{Z_L^{\alpha\beta}}\sum\limits_{j=1}^J A_{J,j} \sum\limits_{k=0}^j \sum\limits_{c=0}^{L-N-J}\sum\limits_{d=0}^{N-J} G_{k,c}(\alpha)G_{j-k,d}(\beta)H_{2J-j,L-N-J-c,N-J-d},
\ee
\end{widetext}
with the combinatorial factors
\begin{eqnarray*}
  G_{k,c}(\alpha)&=&\bin{k+c}{c}\frac{1}{\alpha^{c+k+1}}\\
  H_{K,a,e}&=&\bin{K-1+a+e}{a+e}\bin{a+e}{e}.
\end{eqnarray*}
Through the above we now have the exact form of the sought-after joint probability function for any system size. The form is illustrated in Figure~\ref{fig:Exact}.

\subsection{Thermodynamic limit}
\label{sec:thl}
We here return to (\ref{eq:prob1}). Using the asymptotic form of the normalizing constant given in (\ref{eq:zh}) and (\ref{eq:zl}), we perform a steepest-descent calculation to get the asymptotic results in the large system limit. We consider the different phases individually. Due to the particle-hole symmetry $P_L^{\alpha\beta}(N,J)=P_L^{\beta\, \alpha}(L-N,J)$, it is only necessary to explicitly consider the case $\alpha<\beta$.

\begin{figure}[htp]
  \includegraphics[width=.9\columnwidth]{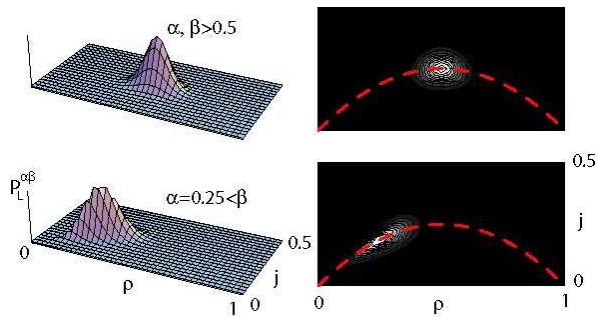}\caption{\label{fig:Asympt} (Color online) The two rows display a surface and a contour plot of the leading behavior of the asymptotic  joint  probability distribution. The calculations were performed at the injection and ejection rates indicated and at a system of size $L=40$ (to make the result comparable to Figure~\ref{fig:Exact}).}
\end{figure}

First turning to the maximal-current phase we use~(\ref{eq:prob1}) together with~(\ref{eq:zh}), and~(\ref{eq:eig}), and drop all pre-factors that are independent of $N$ and $J$ (this will be done throughout), to write
\begin{multline*}
  P^{\alpha\beta}_L(N,J)\sim \frac{4^{J}}{J^{3/2}}\oint\limits_{C_{z},C_{\bar z}} \frac{\d z\d \bar z}{(2\pi\imath)^2}\frac{1}{z^{N-J+1}\bar z^{L-N-J+1}}\\
\times \frac{1}{[1-(z+\bar z)]^{2J-1}}\frac{1}{[2\alpha-1+(z-\bar z)]^2[2\beta-1-(z-\bar z)]^2}.
\end{multline*}
The asymptotic behavior of these integrals is in principle straight forward to calculate. In practice though, it turns out to be quite cumbersome since one has to determine which of the saddle points and lower order poles give the dominant contributions. We can shortcut this through only considering the asymptotic form in some finite region around the peak of the distribution. From the general discussions of the phase diagram in Section~\ref{sec:phase} we know that the average density and current is $\alpha$ and $\beta$ independent. Thus, the lower order poles cannot dictate the asymptotic behavior around the peak value of the probability distribution, and instead this must be set by the saddle points
$$ 
  z^*=\rho-j, \quad \bar z^*=1-\rho-j, \quad \rho= N/L,\quad j= J/L. 
$$
A saddle-point approximation thus results in
\begin{multline}\label{eq:hc}
P^{\alpha\beta}_L(\rho,j)\sim 
\lp\frac{1}{j^{2j}(\rho-j)^{\rho-j}(1-\rho-j)^{1-\rho-j}}\rp^L,
\end{multline}
where we for simplicity have dropped all the sub-dominant pre-factors. Even though the extent of the region of validity of~(\ref{eq:hc}) is unknown, it should be pointed out that the size of this region is a finite fraction of the complete range of $\rho$ and $j$ (as long as the system is away from any phase boundaries). In the first row of Figure~\ref{fig:Asympt} we show the resulting dominating asymptotic plots.

We now turn to the low current, low-density phase. Using~(\ref{eq:prob1}),~(\ref{eq:zl}), and~(\ref{eq:eig}) we have
\begin{multline*}
 P^{\alpha\beta}_L(\rho,j)\sim\oint_{C_{z},C_{\bar z}} \frac{\d z\d \bar z}{(2\pi\imath)^2}\frac{1}{z^{N-J+1}\bar z^{L-N-J+1}}\\
\times\frac{2\alpha-1+z-\bar z}{\beta-\alpha-(z-\bar z)}\frac{1}{(\alpha-\bar z)^{J+1}(1-z-\alpha)^{J+1}}.
\end{multline*}
The same arguments as applied in the high current phase gives the asymptotic probability distribution around the peak. Again it is the saddle points
$$
z^*=\frac{\rho-j}{\rho}(1-\alpha), \quad \bar z^*=\frac{1-\rho-j}{1-\rho}\alpha,
$$
that dominate. The resulting dominant form is
\begin{multline}
\label{eq:lc}
  P^{\alpha\beta}_L(\rho,j)\sim
\\
\lp\frac{\rho^\rho (1-\rho)^{1-\rho}}{\alpha^{1-\rho}(1-\alpha)^{\rho}}\frac{1}{j^{2j}(\rho-j)^{\rho-j}(1-\rho-j)^{1-\rho-j}}\rp^L.
\end{multline}
The above result is directly transferable to the high-density phase through the use of the particle hole symmetry mentioned above.  A realization of the asymptotically dominating part in the low-density phase is shown in the second row of~Figure~\ref{fig:Asympt}. 

It is clear from the asymptotic forms that the probability distribution is non-Gaussian in all phases. This is consistent with the view that long-range correlations are a generic feature of non-equilibrium systems with locally conserved dynamics~\cite{Katz83}.  Sufficiently close to a phase transition, any finite system will reach a point at which the region of validity of the above asymptotic forms shrink to the size of the typical fluctuations. When this happens the system crosses over to a situation where the fluctuations are governed by the tails excluded in the above development. 

It is interesting to note that if we consider the probability density of the current at a fixed density, we recover the same functional form as derived in~\cite{Shaw03} for the probability density of the current in a closed periodic system. 

\subsection{Fluctuations}
It is interesting to note that as the continuum transition is passed, the asymptotic forms~(\ref{eq:hc}) and~(\ref{eq:lc}) indicates that there will be a jump in the density fluctuations of the system. More precisely, in the low-current, low-density phase we have
\begin{multline*}
   C_{{\rm lc,ld}}^{{\rm c}}(\alpha)=\left(\begin{array}{cc}\lv \delta\rho^2\rv&\lv \delta \rho\delta j\rv\\\lv \delta \rho\delta j\rv&\quad \lv\delta j^2\rv\end{array}\right)\\
\sim\frac{\alpha(1-\alpha)}{L}\left(\begin{array}{cc}1  & 1-2\alpha \\ 1-2\alpha & 1-3\alpha(1-\alpha)\end{array}\right).
\end{multline*}
Using the particle-hole symmetry we have
$$
   C_{{\rm lc,hd}}^{{\rm c}}(\beta)=C_{{\rm lc,ld}}^{{\rm c}}(1-\beta),
$$
for the low-current, high-density, phase. In the maximal-current phase we have
$$
  C^{{\rm c}}_{{\rm mc}}\sim\frac{1}{L}\left(\begin{array}{cc}1/8  & 0 \\ 0& 1/16\end{array}\right). 
$$
Thus as we go from either of the low-current phases (regions A or B in the phase diagram in Figure~\ref{fig:pd}) to the maximal current phase (region C in Figure~\ref{fig:pd}), there will be a jump in the correlator
$$
  \Delta C^c\sim-\frac{1}{L}\left(\begin{array}{cc}1/8  & 0 \\ 0& 0\end{array}\right). 
$$
Since we here have a discontinuity in the density fluctuations as we pass over the continuous phase transition, we see that this transition is of second order (c.f. equilibrium statistical mechanics where a discontinuity in the correlator corresponds to a discontinuity in the second-order derivative of the free-energy). The sign of this jump also illustrates that the strength of the fluctuations decrease as we enter the maximal-current phase.
\section{Implications on the KPZ equation}
\label{sec:KPZ}
In this section we briefly point to the known connections between the ASEP and the KPZ equation~\cite{Halpin-Healy95,Stinchcombe01}, and translate our findings to the language of the KPZ equation. 

The ASEP can be mapped onto a lattice growth model~\cite{Halpin-Healy95,Stinchcombe01}, which in turn is believed to share its' long-wavelength characteristics with the KPZ equation. Within the framework of this standard mapping it follows that the growth velocity and the average slope of the interface are respectively given by
$$
  \bar v=j, \quad \overline{\partial h}=1-2\rho.
$$
Letting $\overline{\partial h}|_{{\rm left/right}}$ denote the enforced boundary slopes, we can get the joint slope-velocity distribution for the interface model by substituting$$
\begin{array}{c}
 \rho=(1-\overline{\partial h})/2, \quad j=\bar v \\
\alpha=(1-\overline{\partial h}|_{{\rm left}})/2, \quad \beta=(1+\overline{\partial h}|_{{\rm right}})/2
\end{array}
$$
into the probability distribution for the ASEP. It is further clear that when the average slope at the left boundary, $\overline{\partial h}_{{\rm left}}=1-2\alpha$, matches the slope at the right boundary, $\overline{\partial h}_{{\rm right}}=2\beta-1$, i.e. when $\alpha+\beta=1$, we are at the trivial line of the ASEP where the matrix-product measure reduces to a product measure. Situations where more complicated boundary conditions are relevant have recently been examined experimental and theoretical in~\cite{Myllys03,Ha03}.
\section{conclusion}
In this paper we have examined the joint probability distribution of the system-averaged density and current. We have derived an exact expression for the joint probability function for finite systems, and also considered the thermodynamic limit. This was done by introducing a relaxed operator algebra, and it would be very interesting to examine if the same ``trick'' could somehow be applied to the PASEP. This is  especially important since this model interpolates between a ESS and NESS. The development further shows that even if the normalization constant does not act as a partition function in the normal sense of giving moments through differentiation by a conjugate field, it nevertheless completely governs the fluctuations through~(\ref{eq:prob1}). Sufficiently close to a phase transition, any finite system will reach a point where the above derived asymptotic forms are not valid. Thus it would be interesting to derive the full asymptotic form of the probability distribution, including the tails. We have also shown that the continuous transitions are second order in the sense of equilibrium statistical mechanics, and that it is the density fluctuations that display a discontinuity. Lastly we wrote down the translation of the probability density for the asymptotic ASEP to the probability density of the KPZ equation in terms of the average slope of the interface, and the average interface velocity.
\section{Acknowledgments}
This work was supported by EPSRC under the Oxford Condensed Matter Theory Grant No. GR/R83712/01 and No. GR/M04426. MD gratefully acknowledges support from the Merton College, Oxford, and the Royal Swedish Academy of Science.  
\appendix
\section{The relaxed operator algebra}
\label{app:dual}
The major problem with calculating the probability function as given in~(\ref{eq:obsprob}) is the appearance of the product of the operators
$$
  \left[\frac{1}{1-z\bm  D}\right]_X\bm  D \quad\textrm{and}\quad \left[\frac{1}{1-\bar z\bm  E}\right]_{\bar X}\bm  E
$$
where all the $\bm D$'s are to the left of all the $\bm E$'s. By considering this product we will uncover  operator relations very similar to those of the original operator algebra~(\ref{eq:opalg}). Using the original algebra we can write,
\begin{multline}
\label{eq:fan}
\left[\frac{ 1 }{1-z\bm  D}\right]_X\bm  D\bm  E\left[\frac{1 }{1-\bar z\bm  E}\right]_{\bar X}\\=\left[\frac{1}{1-z\bm  D}\right]_X\bm  D\left[\frac{1}{1-\bar z\bm  E}\right]_{\bar X}\\+\left[\frac{ 1}{1-z\bm  D}\right]_X\bm  E\left[\frac{1}{1-\bar z\bm  E}\right]_{\bar X}.
\end{multline}
From the inversion relation~(\ref{eq:inv}) we have
$$
   \left[\frac{1}{1-\mu \bm  X}\right]_X=1+\mu \bm  X\left[\frac{1}{1-\mu \bm  X}\right]_{X}+\O(\mu^{X+1}),
$$
and using this in the right hand side of~(\ref{eq:fan}) it is easily seen that~(\ref{eq:fan}) is equivalent to
\begin{multline}
  (1-[z+\bar z])\left[\frac{ 1 }{1-z\bm  D}\right]_X\bm  D\left[\frac{1 }{1-\bar z\bm  E}\right]_{\bar X}\bm  E\\
=\left[\frac{ 1 }{1-z\bm  D}\right]_X\bm  D+\left[\frac{1 }{1-\bar z\bm  E}\right]_{\bar X}\bm  E\\
+\O( z^{X+1},\bar z^{\bar X+1}).
\end{multline}
It is now clear that the operators~(\ref{eq:newop}) will satisfy the relation~(\ref{eq:newopalg}). In terms of the operators~(\ref{eq:newop}) the central operator product in~(\ref{eq:obsprob}) reads
\begin{multline*}
  \left[\frac{ 1 }{1-z\bm  D}\right]_{N-1}\bm  D\bm  E\left[\frac{1 }{1-\bar z\bm  E}\right]_{L-N-1}=\frac{\bm  D'\bm  E'}{(1-[z+\bar z])^2}.
\end{multline*}
\section{Generating functional}
\label{app:gen}
By using the definition of the generating functional for $Z^{\alpha\beta}_L$~(\ref{eq:gz}), and the integral representation of the joint probability function~(\ref{eq:prob1}), we can write the generating functional for $N$ and $J$ as
\begin{multline*}
  F^{\alpha\beta}_L(\gamma,\kappa)\Def\sum\limits_{N,J=0}^{\infty}\gamma^N\kappa^J P^{\alpha\beta}_L(N,J)\\
=\frac{\alpha\beta}{Z^{\alpha\beta}_L}\oint_{C_z,C_{\bar z}}\frac{\d z\d \bar z}{(2\pi\imath)^2}\frac{1}{(\bar z-\alpha)(z-\beta)}\frac{1}{\bar z^{L+1}}\frac{1}{z-\gamma\bar z}\\
\times G^{\alpha'\beta'}\lp\frac{z\bar z\kappa}{(1-[z+\bar z])^2}\rp.
\end{multline*}
The contours should be chosen such that the only enclosed poles are the ones at the origins. This object has a rather complicated analytical structure and for simplicity we have restricted our efforts to the probability function.
\section{Exact probability function}
\label{app:exact}
In order to get the exact probability distribution for $N$ and $J$ we need to expand the integrand of~(\ref{eq:prob1}) around the origins of both the $z$- and $\bar z$-plane. This is done in the present section. First we write down the complete expression for the probability function, given the exact form of the normalizing constant as shown in~(\ref{eq:norm})
$$
  Z^{\alpha'\beta'}_L=\sum\limits_{l=1}^LA_{L,l}\sum\limits_{k=0}^{l}\frac{(1-[z+\bar z])^l}{(\alpha-\bar z)^k(\beta-z)^{l-k}}.
$$
Thus we have
\begin{multline*}
   P^{\alpha\beta}_L(N,J)\\
=\frac{\alpha\beta}{Z_L^{\alpha\beta}}\sum\limits_{l=1}^LA_{L,l}\sum\limits_{k=0}^{l}\oint_{C_{z},C_{\bar z}} \frac{\d z\d \bar z}{(2\pi\imath)^2}\frac{1}{z^{N+1-J}\bar z^{L-N+1-J}}\\
\times \frac{1}{(\alpha-\bar z)^{k+1}(\beta-z)^{l-k+1}(1-[z+\bar z])^{2J-l}}.
\end{multline*}
We proceed by considering the expansion factor by factor in their respective Laurent-series around the origins. The first two factors are given by the expansion
$$
  \frac{1}{(\alpha-\bar z)^{k+1}}=\sum\limits_{c=0}^{\infty}G_{k,c}(\alpha)\bar z^c
$$
with
$$
  G_{k,c}(\alpha)=\binom{k+c}{c}\frac{1}{\alpha^{k+c+1}},
$$
while the last factor expands as
$$
  \frac{1}{(1-[z+\bar z])^{2J-l}}=\sum\limits_{a,e=0}^\infty H_{2J-l,a,e}z^e\bar z^a,
$$
with
$$
H_{K,a,e}=\binom{K-1+a+e}{a+e}\binom{a+e}{e}.
$$
Considering the product of all three factors we arrive at
\begin{multline*}
\frac{1}{(\alpha-\bar z)^{k+1}(\beta-z)^{l-k+1}(1-[z+\bar z])^{2J-l}}\\
=\sum\limits_{c,a,d,e=0}^{\infty}G_{k,c}(\alpha)G_{l-k,d}(\beta)H_{2J-l,a,e}\bar z^{c+a} z^{d+e}\\
=\sum\limits_{c,d=0}^{\infty}\sum\limits_{a=c,e=d}^\infty G_{k,c}(\alpha)G_{l-k,d}(\beta)H_{2J-l,a-c,e-d}\bar z^{a} z^{e}
\\=\sum\limits_{a,e=0}^{\infty}\left(\sum\limits_{c,d=0}^{a,e} G_{k,c}(\alpha)G_{l-k,d}(\beta)H_{2J-l,a-c,e-d}\right)\bar z^{a} z^{e}
\end{multline*}
The only terms in this series that will contribute to the probability density are the ones with $a=N-J$ and $e=L-N-J$, and thus we have the exact form of the probability function given by~(\ref{eq:K}).
\bibliography{NewBib2}

\end{document}